\documentstyle[aps,multicol,amsfonts,amssymb,amsmath]{revtex}

\makeatletter
\def\thepage{\@arabic\c@page}
\def\@pnumwidth{2em}
\makeatother

\begin{document}

\draft
\title{\Huge THE VARIANT PRINCIPLE}
\author{N. T. Anh}
\address{Institute for Nuclear Science and Technique,\\
Hanoi, Vietnam. \\
{\footnotesize Email: anh@vaec.vista.gov.vn} }
\date{1999}
\maketitle

\begin{abstract}
Based on the principle of causality, I advance a new principle of variation
and try to use it as the most general principle for research into laws of
nature.
\end{abstract}


\makeatletter
\global\@specialpagefalse
\def\@oddhead{N.T.Anh\hfill The Variant Principle}
\let\@evenhead\@oddhead
\def\@oddfoot{\reset@font\rm\hfill \thepage\hfill
\ifnum\c@page=1
  \llap{\protect\copyright{} 1999 N.T.Anh}%
\fi
} \let\@evenfoot\@oddfoot
\makeatother

\begin{multicols}{2}
\section{INTRODUCTION}

Abstractly, the Nature can be examined as a system of states and actions.
State is a general concept that defines existence, structure, organization,
and conservation of all matter's systems, and that stipulates properties,
inner relationships of all things and phenomena. Action is an operation that
manifests self-influence and inter-influence of states, that presents
dynamic power and impulsion of motion and development. Generally, state is
object on which actions do. Each state has its action. Self-action makes
state conservable and developable. Action of one state on other forms
interaction between them. Self-action and inter-action cause variation of
state from one to other. That variation establishes a general law of motion.

Following this way I advance a new principle -- that is called
Variant Principle. Utilizing this principle as the most general
principle I hope that it is useful for research on a logically
systematic method to review known laws and to predict unknown
laws. And it is a groundwork to unify interactions of nature. I
believe that some of the readers of this article will find out
that this principle explains naturally inner origin of variation,
rules evolutionary processes of things, and perhaps they will be
the ones to complete the quest for theories of the Universe.

The article is organized as follows. In Section 2, I advance the ideas and
concepts for leading the equation of motion. That is just the foundation of
the variant principle. A phenomenon in physics is illustrated by this
principle in Section 3. Conclusion is given in Section 4.

\section{THE EQUATION OF MOTION}

In the Nature, any state and its action are constituent elements of a
subject that I call it \textit{actor},
\begin{equation}
A=({\mathbb{A}}\text{ }\&\text{ }\widehat{\mathrm{A}}),
\tag{I}
\end{equation}
where ${\mathbb{A}}$ is state, and $\widehat{\mathrm{A}}$ is its action
operator.

\begin{enumerate}
\item  For any system in which there is only one \textit{actor} $\{A\}$,
that \textit{actor} is in self-action. This fact causes \textit{actor}
either to be conserved or to be varied by action of itself with respect to
all its possible inner degrees of freedom. Conservation makes \textit{actor}
invariant. But variation obeys an equation of motion,
\begin{equation}
\widehat{\mathrm{A}}{\mathbb{A}}=0,
\tag{II}
\end{equation}
where action operator $\widehat{\mathrm{A}}$ may include differentiation,
integration, and/or other formal operations doing with respect to some
degrees of freedom (such as space, time, and/or some variable), depending on
actually physical problems, and ${\mathbb{A}}$ may naturally be a state
function describing some considered object. The value `$0$' on the right
hand side of Eq. (II) means that variation of \textit{actor} approaches to
stability -- invariance, i.e. self-action is equal to zero when variation
finishes.

Solution of the equation of motion describes variant process of \textit{actor%
}. \textit{Actor} varies and finally becomes to a new \textit{actor}, that is
solution of the equation of motion when variation finishes.

\item  For any system consisting of many \textit{actor}s $\{A_{1};A_{2};...\}
$, each \textit{actor} is in its self-action and actions from others. This
fact causes each \textit{actor} to be varied by actions of itself and others
with respect to all its possible inner and outer degrees of freedom. This
variation obeys an equation of motion,
\begin{equation}
(\widehat{\mathrm{A}}_{1};\widehat{\mathrm{A}}_{2};...)({\mathbb{A}}_{1};%
{\mathbb{A}}_{2};...)=0,
\tag{III}
\label{muleq}
\end{equation}
where action operators $\widehat{\mathrm{A}}_{i}$ of \textit{actor} $A_{i}$
are operations doing with respect to some degrees of freedom, and states $%
{\mathbb{A}}_{i}$ of \textit{actor} $A_{i}$ are functions characterized by
considered objects. The value `$0$' on the right hand side of Eq. (III)
means that actions are equal to zero when variations of \textit{actor}s
finishes, i.e. variations of \textit{actor}s approaches to stability --
invariance. In fact, Eq. (III) is an advanced form of Eq. (II).

Solutions of the equations of motion of \textit{actor}s describe their
variant processes. All \textit{actor}s vary and finally become a new
\textit{actor} $A$, that is solution of the equations of motion when
variations of \textit{actor}s finishes:
\begin{equation}
A=[A_{1},A_{2},...],
\tag{IV}
\end{equation}
where \textit{actor}s are in the same dimension of interaction.

\begin{itemize}
\item[{*}]  For a system consisting of many \textit{actor}s $%
\{A_{1};A_{2};...\}$, the whole system can be considered as a total \textit{%
actor} which includes component \textit{actor}s,
\begin{equation}
\{A\}=\{A_{1};A_{2};...\}.
\tag{V}
\end{equation}
Thereby, \textit{actor} $A$ is in self-action, and it either self-conserves
or self-varies with respect to all its possible inner degrees of freedom.
And variation obeys an equation of motion (II).
\end{itemize}
\end{enumerate}

Hence, the variant principle is stated as follows:

\begin{itemize}
\item[-]  \emph{In the Nature every actor varied by actions of itself and
others with respect to all possible degrees of freedom to become some new
actor is solution of the equation of motion that describes its variant
process.}
\end{itemize}

Indeed, every variation is caused by action of \textit{actor} onto state,
variation is to escape from action, or in other words, state varies to be
agreeable to action. This fact means that under actions \textit{actor} must
vary anyway with respect to all possible degrees of freedom --
transportation facilities to become new \textit{actor}, and that its speed
of variation is dependent on power of action, which is manifested by
conservation of \textit{actor}.

Eigenvalue of action is expressed as instrument to promote variation, as
easiness of variation. Its value over some degree of freedom shows
probability of variation following this direction.

Any \textit{actor} which is done by some action must vary somehow over all
possible degrees of freedom to become new \textit{actor} which is no longer
to be done by any action. That process shows continuous variation of \textit{%
actor} from the beginning to closing.

Therefore, this reality proves that variation is imperative to have its
cause, to have its agent, and that property of variation obeys the equation
of motion.

Thereby, from Eqs. (II) and (III), equation of motion can be built for any
physical law. Using these equations (II) and (III) for research into physics
is considered in the next section. I hope that the readers will understand
more profoundly about the variant principle.

\section{The Rule of Universe's Evolution}

The simplest form of self-action is expansion of actor about some degree of
freedom,
\begin{equation}
e^{\delta x\ \widehat{\partial }_{x}}f(x)=f(x+\delta x). \label{functor}
\end{equation}

Here is just the equation of motion for any quantity $f(x)$, with $x$ degree
of freedom, and $\delta x$ infinitesimal of $x$.

Universe's evolution is described as a law of causality \cite{anh1} essentially
based on just this expansion. The form of Eq.\ (\ref{functor}) is nothing but Taylor's
series. Derivatives of $f(x)$ with respect to $x$ is just variations of $%
f(x) $ over the degree of freedom $x$.

Eq.\ (\ref{functor}) has an important application in modelling the multiplication and the
combination of quanta.

Call $\alpha ,\beta ,\gamma ,...$ quanta. For each quantum there is a rule
of multiplication as follows
\begin{equation}
\alpha ^{n}\rightarrow e^{\partial _{\alpha }}\alpha
^{n}=\sum_{i=0}^{n}C_{i}^{n}\alpha ^{n-i}=(\alpha +1)^{n}  \label{multy}
\end{equation}
where $n$ is order of combination, $\delta \alpha =1$, and $C_{i}^{n}$ is
binary coefficient.

Using Eq.\ (\ref{multy}) I consider two stages in the process of the Universe's
evolution: doublet and triplet.

For two interactive quanta the rule of multiplication reads
\end{multicols}
\rule{8.4cm}{.1mm}\rule{-.1mm}{.1mm}\rule{.1mm}{2mm}
\begin{equation}
\alpha ^{n},\beta ^{n}\rightarrow \frac{1}{2}(e^{\beta \partial _{\alpha
}}\alpha ^{n}+e^{\alpha \partial _{\beta }}\beta
^{n})=\sum_{i=0}^{n}C_{i}^{n}\alpha ^{n-i}\beta ^{i}=(\alpha +\beta )^{n}. \label{doublet}
\end{equation}
And similar to three interactive quanta
\begin{eqnarray}
\alpha ^{n},\beta ^{n},\gamma ^{n}\rightarrow \frac{1}{3}(e^{(\beta +\gamma
)\partial _{\alpha }}\alpha ^{n}+e^{(\gamma +\alpha )\partial _{\beta
}}\beta ^{n}+e^{(\alpha +\beta )\partial _{\gamma }}\gamma ^{n})
&=&\sum_{m}^{n}\sum_{i}^{m}C_{m}^{n}C_{i}^{m}\alpha ^{n-m}\beta
^{m-i}\gamma ^{i}  \nonumber \\
&=&(\alpha +\beta +\gamma )^{n}.  \label{triplet}
\end{eqnarray}
And so fourth. Eqs.\ (\ref{doublet}) and (\ref{triplet}) can be drawn as schemata.

\begin{equation}
\begin{array}{cccccccccccccc}
\vdots &  &  &  & \cdots &  & \cdots &  & \cdots &  & \cdots &  &  &  \\
&  &  &  &  &  &  &  &  &  &  &  &  &  \\
\overline{2} &  &  &  &  &  & \overline{1} &  & \overline{1} &  &  &  &  &
\\
&  &  &  &  &  &  &  &  &  &  &  &  &  \\
0 &  &  &  &  &  &  & \bigcirc &  &  &  &  &  &  \\
&  &  &  &  &  &  &  &  &  &  &  &  &  \\
\underline{2} &  &  &  &  &  & 1 &  & 1 &  &  &  &  &  \\
&  &  &  &  &  &  &  &  &  &  &  &  &  \\
\underline{2}\otimes \underline{2}=\underline{3}\oplus \underline{1} &  &  &
&  & 1 &  & 2 &  & 1 &  &  &  &  \\
&  &  &  &  &  &  &  &  &  &  &  &  &  \\
\underline{2}\otimes \underline{2}\otimes \underline{2}=\underline{4}\oplus
\underline{2}\oplus \underline{2} &  &  &  & 1 &  & 3 &  & 3 &  & 1 &  &  &
\\
&  &  &  &  &  &  &  &  &  &  &  &  &  \\
\vdots &  &  & 1 &  & 4 &  & 6 &  & 4 &  & 1 &  &  \\
&  &  &  &  &  &  &  &  &  &  &  &  &  \\
\vdots & \cdots &  & \cdots &  & \cdots &  & \cdots &  & \cdots &  & \cdots
&  & \cdots
\end{array} \label{pastri}
\end{equation}
is the schema for Eq.\ (\ref{pastri}), where $\underline{2}$ means two quanta $\alpha $
and $\beta $. The numbers in the triangle is the binary coefficients which
are called weights of classes. For example,
\begin{equation*}
\underline{2}\otimes \underline{2}=\underline{3}\oplus \underline{1}=
\begin{array}{ccccc}
&  & 1 &  &  \\
1 & \text{--------} & 1 & \text{--------} & 1
\end{array}
.
\end{equation*}
And similar to Eq.\ (\ref{triplet}) it reads

\begin{equation}
\begin{array}{cccccccccccccc}
\vdots  &  &  &  & \overline{1} &  &  &  &  &  &  &  &  &  \\
\overline{3} &  &  & \overline{1} &  &  & \overline{1} &  &  &  &  &  &  &
\\
&  &  &  &  &  &  &  &  &  &  &  &  &  \\
&  &  &  &  &  &  &  &  &  &  &  &  &  \\
0 &  &  &  &  & \bigcirc  &  &  &  &  &  &  &  &  \\
&  &  &  &  &  &  &  &  &  &  &  &  &  \\
&  &  &  &  &  &  &  &  &  &  &  &  &  \\
3 &  &  &  & \mathbf{1} &  &  & \mathbf{1} &  &  &  &  &  &  \\
&  &  &  &  &  & \mathbf{1} &  &  &  &  &  &  &  \\
&  &  &  &  &  &  &  &  &  &  &  &  &  \\
&  &  &  &  &  &  &  &  &  &  &  &  &  \\
&  &  & \mathbf{1} &  &  & 2 &  &  & \mathbf{1} &  &  &  &  \\
\underline{3}\otimes \underline{3}=\underline{6}\oplus \overline{3} &  &  &
&  & \mathbf{2} &  &  & \mathbf{2} &  &  &  &  &  \\
&  &  &  &  &  &  & \mathbf{1} &  &  &  &  &  &  \\
&  &  &  &  &  &  &  &  &  &  &  &  &  \\
&  & \mathbf{1} &  &  & 3 &  &  & 3 &  &  & \mathbf{1} &  &  \\
\underline{3}\otimes \underline{3}\otimes \underline{3}=\underline{10}\oplus
8\oplus 8\oplus 1 &  &  &  & \mathbf{3} &  &  & 6 &  &  & \mathbf{3} &  &  &
\\
&  &  &  &  &  & \mathbf{3} &  &  & \mathbf{3} &  &  &  &  \\
&  &  &  &  &  &  &  & \mathbf{1} &  &  &  &  &  \\
&  &  &  &  &  &  &  &  &  &  &  &  &  \\
& \mathbf{1} &  &  & 4 &  &  & 6 &  &  & 4 &  &  & \mathbf{1} \\
&  &  & \mathbf{4} &  &  & 12 &  &  & 12 &  &  & \mathbf{4} &  \\
\underline{3}\otimes \underline{3}\otimes \underline{3}\otimes \underline{3}
&  &  &  &  & \mathbf{6} &  &  & 12 &  &  & \mathbf{6} &  &  \\
&  &  &  &  &  &  & \mathbf{4} &  &  & \mathbf{4} &  &  &  \\
\vdots  &  &  &  &  &  &  &  &  & \mathbf{1} &  &  &  &
\end{array}
\end{equation}
where $\underline{3}$ means three quanta $\alpha $, $\beta $ and $\gamma $.
The coefficients in the pyramid are weights of classes,
\begin{equation*}
\begin{tabular}{lllllllll}
&  &  &  & $1$ &  &  &  &  \\
&  &  & $\mathbf{1}$ &  &  & $\mathbf{1}$ &  &  \\
& - & - & - & - & - & - & - & - \\
$\underline{3}\otimes \underline{3}=\underline{6}\oplus \overline{3}=$ & $%
\mathbf{1}$ &  &  & $1$ &  &  & $\mathbf{1}$ &  \\
&  &  & $\mathbf{1}$ &  &  & $\mathbf{1}$ &  &  \\
&  &  &  &  & $\mathbf{1}$ &  &  &
\end{tabular}
,
\end{equation*}
\begin{equation*}
\begin{tabular}{llllllll}
&  &  &  & $1$ &  &  &  \\
& - & - & - & - & - & - & - \\
$\underline{3}\otimes \overline{3}=1\oplus 8=$ &  & $1$ &  &  & $1$ &  &  \\
& $\mathbf{1}$ &  &  & $2$ &  &  & $\mathbf{1}$ \\
&  &  & $\mathbf{1}$ &  &  & $\mathbf{1}$ &
\end{tabular}
.
\end{equation*}
\hspace{9cm}\rule{.1mm}{-4mm}\rule{.1mm}{.1mm}\rule{8cm}{.1mm}
\begin{multicols}{2}
It is easily to identify that the above schemata have the forms similar to
the $SU(2)$ and the $SU(3)$ groups. This means that for $n$ quanta there is
a corresponding schema according to the $SU(n)$ group, and the
multiplication and the combination of the Universe conform to the $SU$
group. This rule is studied further in Ref. \cite{anh2}.

\section{CONCLUSION}

The theory of causality \cite{chi} is very useful to understand about the cause of
variation. The coexistence of two different actors causes a contradiction.
The solution to contradiction makes contradiction varied. That variation is
just one of each actor inclining to become a new actor. It means the
difference and the contradiction of two actors have inclining towards zero.
Indeed, every system comes to equilibrium, stability. A some state which has
any immanent contradiction must vary to become a new one having no
contradiction.

The variant principle deals with the law of variation of actors, describes
only actors with their actions and states, not to mention the difference and
even the contradiction in them. In insight the variant principle is more
elementary and easier to understand than the causal principle since
everything is referred as actor existing in nature. Self-action and
inter-action of actors onto their states cause the world to be in motion and
in variation.

Although the variant principle gives a powerful fundamental for application
to research into laws of nature, there is no rule arisen yet for formulizing
self-action and inter-action operators. However, there are some ways to
enter operators in the equation of motion that I hope that in some next
article this ways will be synthesized to a standard rule.

For instance, in the quantum electromagnetic dynamics the equations of
motion of the electron-positron and the electromagnetic field are:
\begin{eqnarray*}
i\gamma ^{\mu }\partial _{\mu }\psi (x)+\frac{m_{e}c}{\hbar }\psi (x)+\frac{e%
}{\hbar }\gamma ^{\mu }A_{\mu }(x)\psi (x) &=&0, \\
\square A_{\mu }+ie\overline{\psi }(x)\gamma _{\mu }\psi (x) &=&0.
\end{eqnarray*}
The first line is the equation of motion of electron, the first term
corresponds to the variation of electron with respect to space-time, the
second gives conservation of electron, and the third is action of the
electromagnetic field onto electron. The second line can be rewritten as
\begin{equation*}
\partial ^{\nu }F_{\nu \mu }-J_{\mu }=0,
\end{equation*}
that is nothing but the Maxwell equation, with $F_{\nu \mu }=\partial _{\nu
}A_{\mu }-\partial _{\mu }A_{\nu }$ the electromagnetic field tenser, $%
A_{\mu }$ the 4-dimensional potential, $J_{\mu }=-ie\overline{\psi }%
(x)\gamma _{\mu }\psi (x)$ the 4-dimensional current density, the first term
corresponds to the variation of the electromagnetic field, the second is the
external current density of the electromagnetic field, (here the mass of
photon is zero, so the mass term is not present).

This example is easy to show that:

\begin{itemize}
\item[--]  The variation done over some degree of freedom is expressed as
derivation with respect to that degree of freedom.

\item[--]  The conservation of actor is written as a term of actor
multiplied by a constant characterized by its conservation.

\item[--]  The influence of other actor on an actor is represented as a
multiplication of two actors.

\item[--]  The external actor stands equally with its variation, when an
external influence does on an actor as an external current, an external
source, or an external force.
\end{itemize}

In conclusion, it is the fact that some readers think that the
variant principle is rather in philosophy than in physics. That is
not true. Doing physics is discovery of nature, not only matter
composition, phenomena, processes, but also more necessary, laws,
principles, and rules. In reality, principia are profound elements
of physics, and of course they have something closing to
philosophy. The variant principle is one of principia for research
on nature. And I hope that discovering principia will be one of
new directions to do physics.

\section*{Acknowledgments}

We would like to thank Dr. D. M. Chi for useful discussions and valuable
comments.

The present article was supported in part by the Advanced Research Project
on Natural Sciences of the MT\&A Center.

\end{multicols}
\end{document}